\AtBeginDocument{\date{15 August 2018}}

\documentclass[prd,preprintnumbers,nofootinbib,twocolumn,longbibliography]{revtex4-1}
\usepackage{fix-revtex-bib}  

\usepackage{graphicx}
\usepackage{amsmath}
\usepackage{amssymb}
\usepackage{braket}
\usepackage{color}
\usepackage{units}
\usepackage{soul}
\usepackage[usenames,dvipsnames]{xcolor}
\usepackage[normalem]{ulem}
\usepackage[bookmarksopen,bookmarksnumbered]{hyperref}

\newcommand\3[1]{\boldsymbol{#1}}

\begin{document}

\title{Graphical Structure of Hadronization and Factorization in Hard Collisions}

\preprint{JLAB-THY-17-2610}    
\author{John Collins}
\email{jcc8@psu.edu}
\affiliation{%
  Department of Physics, Penn State University, University Park PA 16802, USA}
\author{Ted C. Rogers}
\email{trogers@odu.edu}
\affiliation{Theory Center, Jefferson Lab, 12000 Jefferson Avenue, Newport News, 
VA 23606, USA}
\affiliation{Department of Physics, Old Dominion University, Norfolk, VA 23529, 
USA}  

\begin{abstract}
  Models of hadronization of hard jets in QCD are often presented in
  terms of Feynman-graph structures that can be thought of as
  effective field theory approximations to dynamical non-perturbative
  physics in QCD.  Such models can be formulated as a kind of
  multiperipheral model.  We obtain general constraints on such models
  in order for them to be self-consistent, and we relate the
  constraints to the space-time structure of hadronization.  We show
  that appropriate models can be considered as implementing
  string-like hadronization.  When the models are put in a
  multiperipheral form, the effective vertices and/or lines must be
  momentum non-conserving: they take 4-momentum from the external
  string-like field.
\end{abstract}

\pacs{PACS??}

\maketitle

\section{Introduction}
\label{sec:intro}

An important topic in QCD is to properly understand the interface
between non-perturbative hadronization and factorization physics (with
its perturbative content). This is especially important with the
current widespread interest on the details of partonic interactions in
hadronic and nuclear physics.  An immediate motivation for the work
described in this paper is the need for incorporating non-perturbative
polarization effects in Monte-Carlo event generators.  (See
Ref.~\cite{Sjostrand:2016bif} for an up-to-date overview of MCEGs.)

The polarization effects at issue are those responsible for the much
studied Sivers and Collins functions and related quantities.  The
primary complications concerning event generators arise because event
generators are formulated in terms of probabilistic processes for the
different components of a reaction.  In contrast, interesting
polarization effects involve quantum-mechanical entanglement between
different parts of the reaction.  A simple example is given by the
azimuthal correlation between back-to-back pairs of hadrons in
$e^+e^-$ annihilation \cite{Artru:1995zu}, where the correlation (via
polarized dihadron fragmentation functions for a primary
quark-antiquark pair) arises because the quark and antiquark are in an
entangled spin state.  Such entanglement implies that the
fragmentation of the two jets is not literally independent even if the
cross section is written as the product of two fragmentation
functions.

Work proposing implementations of non-perturbative polarization
effects in event generators is found in \cite{Kerbizi:2017rpp,
  Matevosyan:2016fwi, Bentz:2016rav, Kerbizi:2018qpp}.  To treat
polarization in a way that is fully consistent with the underlying
principles of quantum theory, the formulations are made in terms of
Feynman graph structures.  These structures can be treated as
manifestations of an effective field theory that usefully approximates
the true non-perturbative behavior of QCD in the relevant kinematic
regime.  One notable feature of this framework is the determination of
the most general polarization structure in the elementary splitting
functions \cite{Matevosyan:2016fwi}.

The work in \citet{Kerbizi:2017rpp} provides in a preliminary version
a phenomenologically successful Monte-Carlo simulation of the
hadronization of the jet produced by a transversely polarized quark.
Of the model's 5 free parameters, 4 concern unpolarized fragmentation,
and were fitted to unpolarized data in semi-inclusive deeply inelastic
lepton-hadron scattering (SIDIS).  The single remaining parameter
needed for the polarized process was obtained from data on the Collins
asymmetry in $e^+e^-$ annihilation to hadrons. The model then
successfully predicted the Collins asymmetry in SIDIS data from the
COMPASS experiment \cite{Adolph:2014zba}, with 18 data points, as well
as the dihadron asymmetry in the same experiment. 
The model was based on a string-model formulation by Artru and
Belghobsi \cite{Artru:2012zz, Artru:2010st, Artru:in2p3-00953539}.
While the success is encouraging, the method is yet to be incorporated
in a full event generator.  In addition, the effects of primary vector
meson production were not incorporated, so the non-perturbative
mechanisms implemented are not the whole story.

Now factorization theorems do incorporate non-perturbative effects in
the form of parton densities and fragmentation functions.  However, as
one of us explained in \cite{Collins:2016ztc}, there is a mismatch
between the measured properties of hadronization of hard jets and the
order-by-order asymptotic properties used in existing factorization
proofs.  In the asymptotics of individual graphs, there arises strong
ordering of kinematics between different parts of the graphs, and the
resulting large rapidity differences are used in an essential way in
the proofs, notably to factor out the effects of soft gluon subgraphs
from collinear subgraphs.  But in reality, hadronization gives rise to
approximately uniform distributions in rapidity with no large gaps.
Event generators need to model such event structures.

Related to this is the following conceptual mismatch.  Intuitively one
explains the process in a space-time picture.  There is first a hard
collision in which a state of some number of partons is generated
over a small distance and time scale.  Only at distinctly later times
does the partonic state turn into observed hadrons.  In contrast,
factorization proofs as well as actual calculations are made in terms
of ordinary momentum-space Feynman graphs.

The primary purpose of this paper is therefore to provide an analysis
of the structures in models of hadronization that are presented in
Feynman-graph form, such as those found in or implied by
\cite{Kerbizi:2017rpp, Matevosyan:2016fwi, Bentz:2016rav,
  Kerbizi:2018qpp}, and to determine which kinds of model are
appropriate and which are not.  We will concentrate exclusively on the
application to $e^+e^-$ annihilation to hadrons in the 2 jet case,
although the principles are more general.

There are two areas that form essential background to our treatment.

\begin{figure}
  \centering
  \includegraphics[scale=0.5]{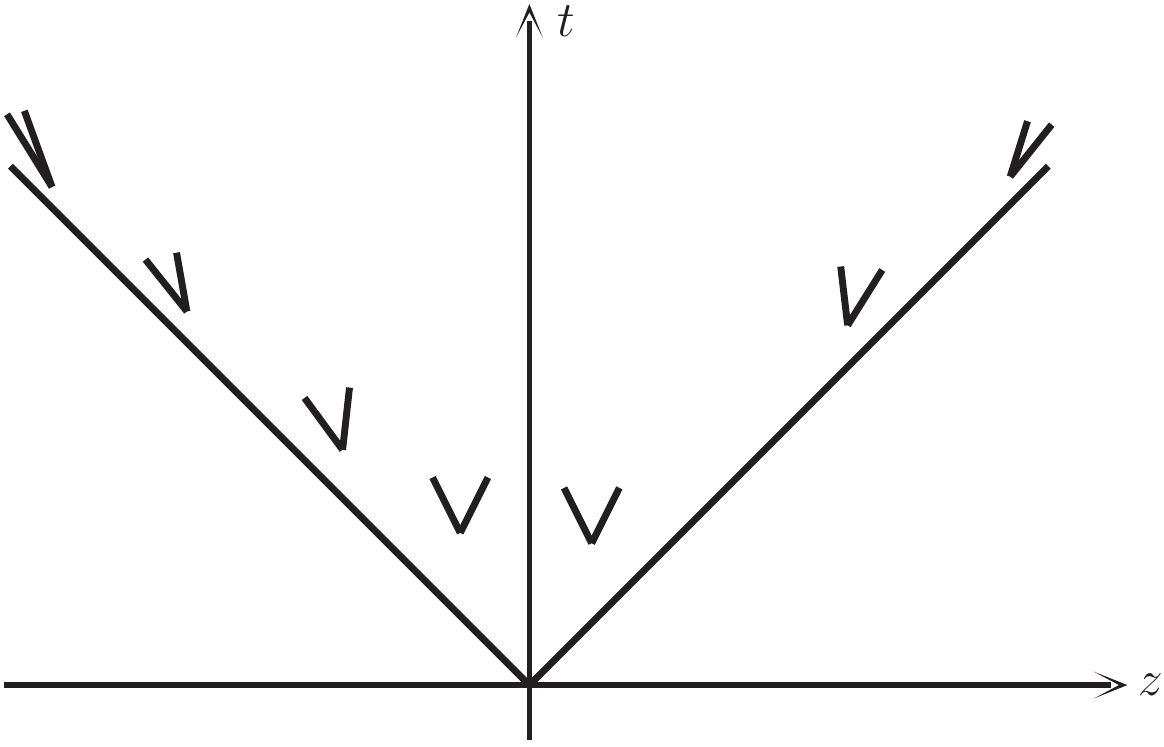}
  \caption{String model \cite{Artru:1979ye, Andersson:1983ia} for
    hadronization of quark-antiquark pair in $e^+e^-$ annihilation,
    pictured in space-time. A string (or flux tube) is created between
    the outgoing quark and antiquark. Quark-antiquark pairs are
    created in the color field in the flux tube, and then combine into
    color singlet hadrons.  The formation of the pairs occurs
    dominantly near a space-like hyperbola an invariant distance of
    order 1 $\textrm{fm}/c$ to the future of the approximately
    trajectories of the original quark and antiquark.}
  \label{fig:string}
\end{figure}

The first concerns standard string models of hadronization models such
as were introduced long ago by Artru and Mennessier
\cite{Artru:1974hr} and by Field and Feynman \cite{Field:1977fa}.  A
particularly attractive qualitative description in the context of QCD
was given in \cite{Field:1977fa} for the case of two-jet production in
$e^+e^-$ annihilation: The high-energy outgoing quark and antiquark
form a tube or string of color flux between them.  Quark-antiquark
pairs are generated in the strong color field, and then reassemble
themselves into color-singlet hadrons.  A detailed quantitative
dynamical description in semi-classical form in space-time was
provided in \cite{Artru:1974hr}.  Later elaborations
\cite{Andersson:1983ia, Artru:1979ye, Andersson:1998tv} led to the
Lund string model, used in the PYTHIA event generator
\cite{Sjostrand:2006za, Sjostrand:2007gs}.  Closely related are
cluster models \cite{Field:1982dg, Webber:1983if} of hadronization.
See Fig.\ \ref{fig:string} for the space-time structure.

General features of string models include the chain decay ansatz,
independent fragmentation, and an iterative scheme for generating
final state hadrons \cite{Field:1977fa}.  However, these classic
hadronization models leave certain interesting issues, notably
hadronization of polarized partons, unaddressed.  This has led to
recent proposals for extending hadronization models to include spin
effects~\cite{Kerbizi:2017rpp, Matevosyan:2016fwi, Bentz:2016rav,
  Kerbizi:2018qpp}.

A second background area to our work stems from the experimental
observation that the particles observed in the final states of jets
are approximately uniformly distributed in rapidity, e.g.,
\cite{Adolph:2014zba}, with the number of hadrons per unit rapidity
depending weakly on the high-energy scale $Q$ or $\sqrt{s}$.

\begin{figure}
  \centering
  \includegraphics[scale=0.5]{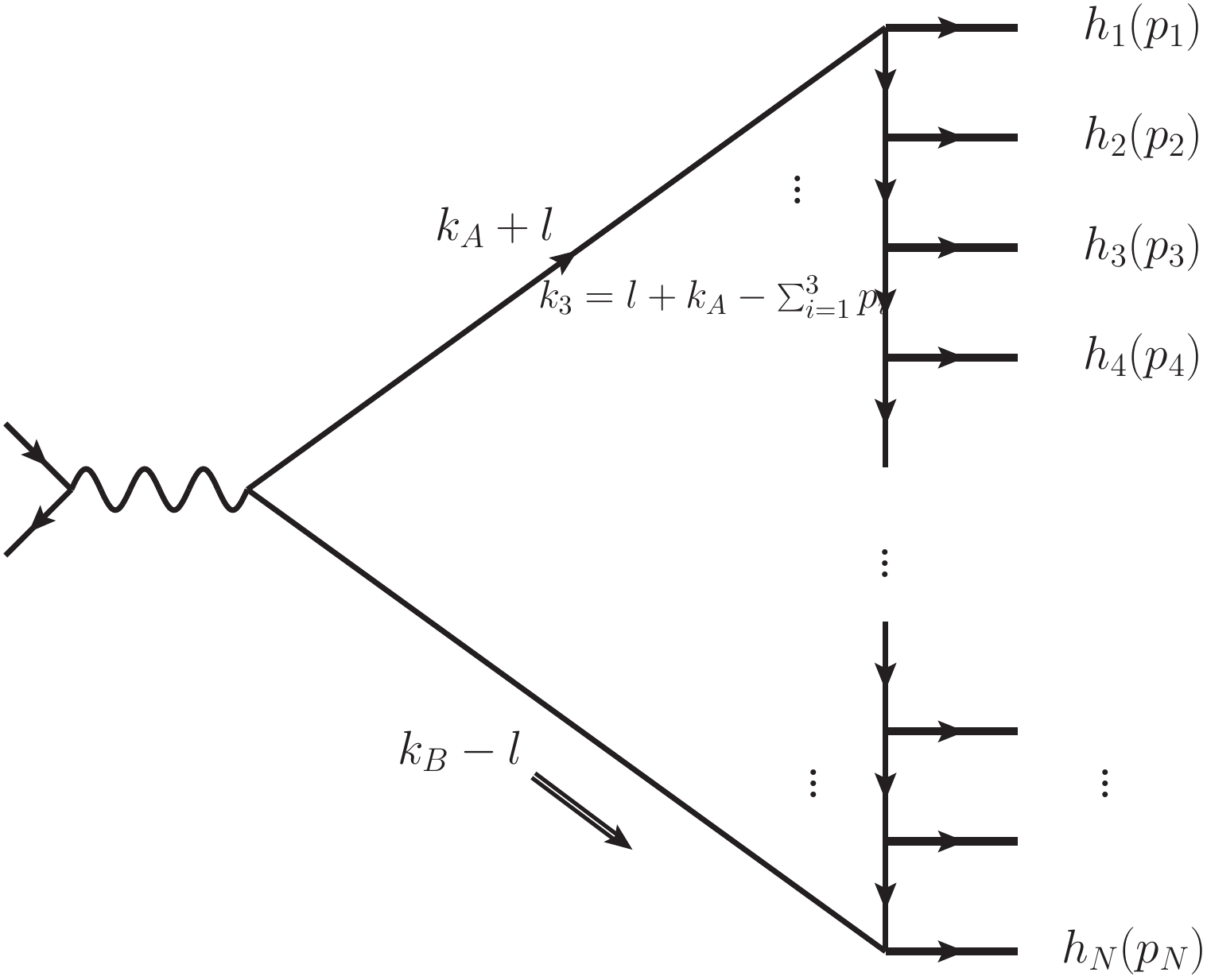}
  \caption{Elementary multiperipheral model for hadronization of
    quark-antiquark pair in $e^+e^-$ annihilation.}
  \label{fig:mpm}
\end{figure}

In the case of soft/minimum-bias hadron-hadron collisions, a simple
and natural initial candidate model in Feynman-graph form is a
multiperipheral model (MPM)~\cite{ELOP}.  Applied to $e^+e^-$
annihilation, using the structure of the MPM would give Fig.\
\ref{fig:mpm}. Here, a quark-antiquark pair is generated from a
virtual photon; the quark and antiquark go outwards.  In the Feynman
graph, the quark line is formed into a loop, and the hadrons are
generated along that line, at momentum-conserving quark-hadron-quark
vertices (which should be effective vertices in QCD).  Because the
line exchanged in the vertical channel is of a spin-half field, a
single graph of fixed order is power suppressed as $Q$ increases.  But
the typical number of particles produced is intended to increase with
$Q$ proportionally to $\ln(Q^2/m^2)$, and in a strong-coupling
situation this allows the result to be unsuppressed.  The core idea of
the MPM is a hypothesis \cite{Kogut:1973fn} that the relevant
interactions only occur between quanta with nearby kinematics.

We emphasize the MPM because it corresponds to graphical structures
that appear to be used in the work on polarized event generators
\cite{Kerbizi:2017rpp, Matevosyan:2016fwi, Bentz:2016rav,
  Kerbizi:2018qpp}.  In the case of the work by Artru, Kerbizi and
collaborators \cite{Artru:2010st, Artru:in2p3-00953539, Artru:2012zz,
  Kerbizi:2018qpp}, MPM graphs are explicitly given.  In the case of
Matevosyan et al.\ \cite{Matevosyan:2016fwi}, the connection is less
direct.  They present their work as involving a sequence of splitting
vertices, and in earlier work from the same group,
\cite{Bentz:2016rav} and especially \cite{Ito:2009zc}, the splitting
vertices are treated as vertices in an effective theory.  Assembling
the vertices to make a model for the amplitude for the whole process
gives MPM graphs.  In both cases, these appearances are misleading.

\begin{figure}
  \includegraphics[scale=0.5]{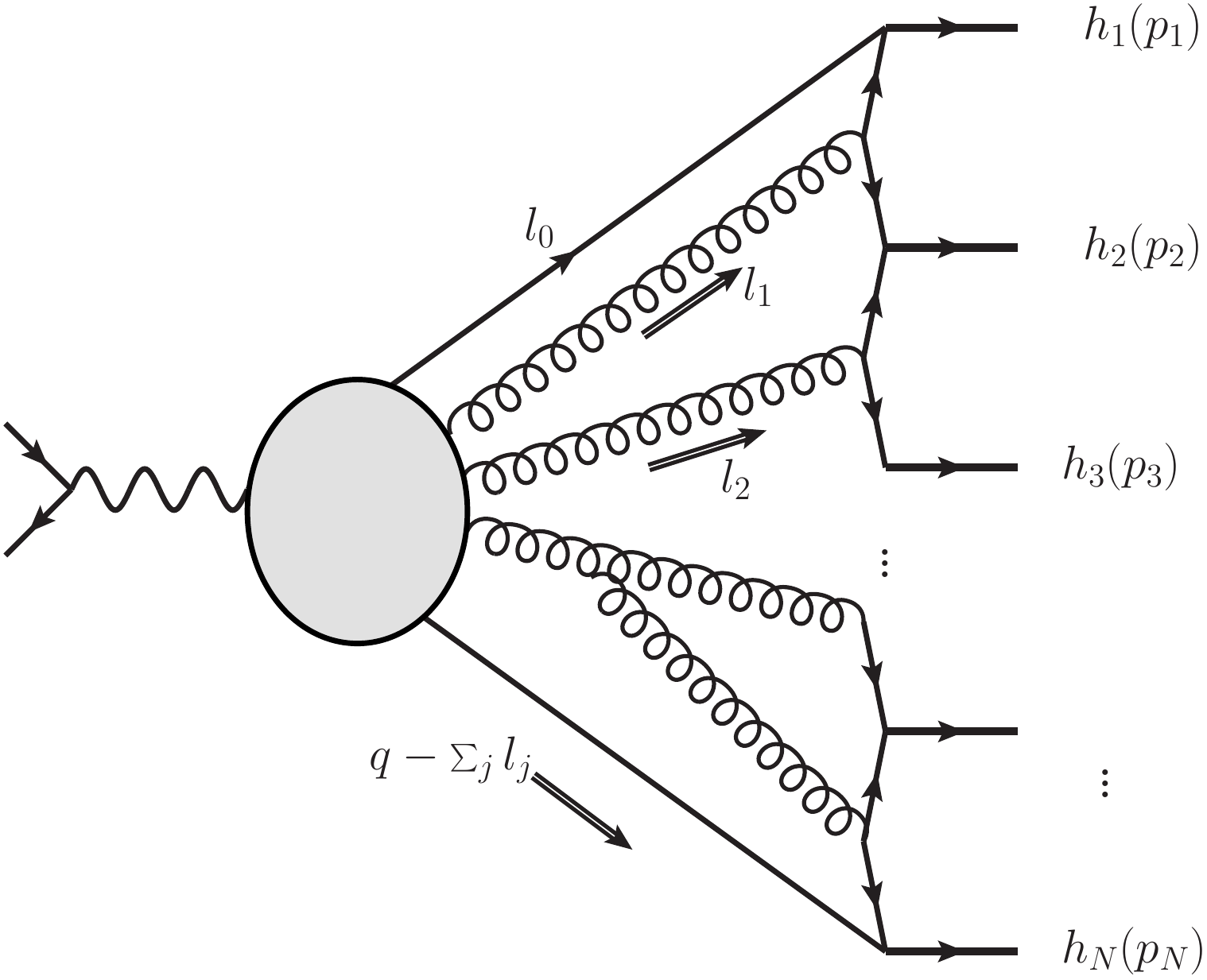}
  \caption{Modified multiperipheral model. It is modified from Fig.\
    \ref{fig:mpm} by allowance for the emission of gluons.  Note that
    the gluons may arise not only directly from the initial quark and
    antiquark lines but also, for example, from a splitting, as
    shown.}
  \label{fig:mpm2}
\end{figure}

In this paper, we first show that taking the MPM, Fig.\ \ref{fig:mpm},
literally does not work.  We show that the loop momentum can be
deformed out of the intended region of kinematics into a region where
short-distance pQCD physics is valid and the model, with its effective
non-perturbative vertices is invalid.  We then show that a motivated,
minimal modification that does work is Fig.\ \ref{fig:mpm2} with extra
gluon lines; this matches both cluster hadronization and models of the
Lund-string type, but now in Feynman-graph form.  We will point out
that models of this form could be considered a MPM, but with the
quark-quark-hadron vertices and the associated connecting quark lines
no longer being momentum conserving.  Instead they should be
considered as absorbing energy and momentum from the color flux tube.
This dramatically changes the space-time structure.

Consequently, the interpretation of apparently MPM-like structures in
work on hadronization like Refs.\ \cite{Kerbizi:2017rpp,
  Matevosyan:2016fwi, Bentz:2016rav, Kerbizi:2018qpp} must be modified
to match string physics.  Careful examination of these papers shows
that this is indeed the case: Implementation of the kinematics of
multiple splitting involves momentum non-conservation at the level of
MPM graphs that corresponds to implementations of a string model.

Many elements of our work can be found in the literature, e.g., a
graph like Fig.\ \ref{fig:mpm2} for cluster hadronization.  Just
before QCD was formulated, Kogut, Sinclair, and Susskind
\cite{Kogut:1973fn} showed that multi-peripheral models cannot be
applicable to hadronization in deep-inelastic scattering and $e^+e^-$
annihilation. Shortly afterwards, Casher, Kogut, and Susskind
\cite{Casher:1973uf, Casher:1974vf}, still before the advent of QCD,
proposed what can now be called a string model.  Our argument is a
modernized version of those old arguments taking account of the much
better knowledge we now have of QCD, and putting it in a new context.

The importance of formulating a hadronization model in terms of
Feynman graphs (with effective vertices and propagators) is that it
automatically obeys general principles of quantum mechanics and
quantum field theory, of causality and of Lorentz invariance.  They
are natural arenas for consistently incorporating spin effects,
especially with entangled spin states, which are hard to formulate
purely semi-classically.  

There is interesting work on effective field theory for chiral
symmetry breaking \cite{Schweitzer:2012hh}.  A better understanding of
the imperatives of Feynman graph implementations of hadronization
should lead to suggestions as to appropriate ways of applying the
chiral models to hadronization.

\section{The elementary MPM}
\label{sec:mpm}

\subsection{Basic Setup}
\label{sec:basic}

Consider the graph shown in Fig.\ \ref{fig:mpm}, as a possible model
for production of hadronic final states in $e^+e^-$ annihilation.  In
accordance with data, we assume that a small number of particles (around
three pions\footnote{This estimate can be roughly deduced from
  measurements by the TASSO collaboration \cite{Braunschweig:1990yd}
  --- see App.\ \ref{sec:TASSO}.}) is produced per unit rapidity, and with
a limited transverse momentum all with respect to a jet axis (e.g.,
the thrust axis).  The typical transverse momentum is perhaps 0.3 or
0.4 GeV.

We will work in a center-of-mass frame, and label the particles in
order of rapidity along the jet axis.  To define the $z$-axis, we
choose hadron 1 to have zero transverse momentum and positive
rapidity.  Let the center-of-mass energy be $Q$.  In light-front
coordinates $(+,-,T)$, the virtual photon's momentum is $q =
(Q/\sqrt2,Q/\sqrt2,0_T)$.  Then we write the hadron momenta in terms
of rapidity and transverse momentum:
\begin{equation}
  \label{eq:pj}
  p_j = \left( \frac{E_{j,T}}{\sqrt2} e^{y_j}, \frac{E_{j,T}}{\sqrt2} e^{-y_j}, \3{p}_{j,T}\right),
\end{equation}
where $E_{j,T} = \sqrt{p_{j,T}^2 + m_h^2}$ and $m_h$ is the mass of
the hadron.  We have chosen $\3{p}_{1,T}=0$.  The momenta must obey
momentum conservation: $\sum_{j=1}^N p_j = q$.

Given the assumptions of the model, a rough estimate of the hadron
rapidities is given by
\begin{equation}
  \label{eq:yj}
  y_j \sim \frac{(N+1)-2j}{N-1} \ln \frac{Q}{m}, 
\end{equation}
with the number of produced particles $N$ being approximately
proportional to $\ln(Q^2/m^2)$, with a coefficient of around 3,
as already mentioned.  (The coefficient
could be rather less if the primary hadrons are mostly vector mesons
instead of pions.)

For the purposes of the model, we assume that the internal loop line
of Fig.\ \ref{fig:mpm} is for a quark, and that the hadrons are pions.
But we will not use that assumption in any detail, since our concern
is only with analytic properties, as determined by propagator
denominators.

We define the origin of the loop momentum $l$ by writing the quark
momenta from the electromagnetic vertex as $k_A+l$ and $k_B-l$, and
defining
\begin{equation}
  \label{eq:kA.kB}
  k_A = (Q/\sqrt2, 0, \3{0}_T ),
\qquad
  k_B = (0, Q/\sqrt2, \3{0}_T ),
\end{equation}
which would be the quark and antiquark momenta if they were free and
massless.  Thus $l$ parameterizes the deviation of quark kinematics
from parton-model values.

The momentum $k_j$ of the line between hadron $j$ and hadron $j+1$ is
\begin{align}
  k_j ={}& l + k_A - \sum_{i=1}^j p_i
\nonumber\\
    ={}& l- k_B + \sum_{i=j+1}^N p_i .
\end{align}
Hence
\begin{equation}
  \label{eq:kj+}
  k_j^+ = l^+ + \sum_{i=j+1}^N p_i^+
     ~ = ~ l^+ + \Theta(m e^{y_j})
\end{equation}
and 
\begin{equation}
  \label{eq:kj-}
  k_j^- = l^- - \sum_{i=1}^j p_i^-
     ~ = ~ l^- - \Theta(m e^{-y_j}).
\end{equation}
In the last term in each equation, estimates are given with the aid of
Knuth's notation \cite{Knuth:asymptotic} $\Theta(\dots)$ rather than the
conventional order notation $O(\dots)$ to emphasize that the estimates
are to within a finite factor. The usual ``big $O$" notation would allow the
actual result to be arbitrarily much smaller, which is not the case
here. The estimates arise as follows: Because of the approximately
uniform distribution of hadrons in rapidity, the sum in the equations
is of an approximately geometrical series.  Then the sums are
dominated by the few terms whose index $i$ is nearest to $j$.

In analyzing the properties of the graph, we find it useful to
  consider first a baseline case given by $l=0$. The power counting
for all lines, i.e., the sizes of their propagator
  denominators is then established by
the above estimates.  Each $k_j^+$ is strictly positive
and each $k_j^-$ is strictly negative.  This implies that the hadrons get
their (mostly large) plus-momentum component from the upper quark
line, and their minus momentum from the lower line.  These conditions,
and the sizes calculated, ensure that the virtualities of all the
lines are of order $\Theta(m^2)$, as is appropriate for a model of
non-perturbative physics in QCD.

\subsection{Integral over loop momentum $l$}

Within the above description, the virtualities in the graph in
Fig.\ \ref{fig:mpm} appear to be consistent with what one
  might expect for a model of non-perturbative physics, at least
if $l=0$. However, $l$ is an integration variable. 

Consider the process as having evolution in space-time from a
quark-antiquark pair at short distances to a hadronic state at large
distance.  Then the conversion to hadrons occurs with unit
probability. So the power law for making the quark-antiquark pair is
the same as power counting for the full $e^+ e^- \to \text{hadrons}$
cross section. That is, hadronization is a leading power effect.  This
also implies that the size of the non-perturbative effective
quark-quark-hadron vertex must be such as to self-consistently give
exactly the unit probability of hadronization.

Furthermore, for the MPM to implement the intended non-perturbative
physics, $k_A + l$ and $k_B - l$ must have a normal non-perturbative
virtuality,
\begin{equation}
(k_A+l)^2 = \Theta(m^2) \, , \qquad (k_B-l)^2 = \Theta(m^2) \, .
\end{equation}
This also indicates that hadronization occurs long after the hard
vertex.  

If the integral were to stay in the region relevant for the assumed
non-perturbative physics, then the transverse components of $l$ would be $\Theta(m)$,
while the longitudinal components would be 
$\Theta(m^2/Q)$, i.e., $l$ would be in the Glauber region:  
\begin{equation}
l = \left( \Theta\left( \frac{m^2}{Q} \right) , \Theta\left(\frac{m^2}{Q} \right), \Theta(\3{m}_T) \right) \, . 
\end{equation}
This is simply because the virtualities of the $k_A + l$ and
  $k_B - l$ lines are:
\begin{equation}
(k_A + l)^2 = 2 k_A^+ l^- + l^2\, , \qquad (k_B - l)^2 = -2 k_B^- l^+ + l^2 \, ,
\end{equation}
and $k_A^+$ and $k_B^-$ are of order $Q$.
If, however, $l$ is merely normal-soft (i.e., $l = (O(m), O(m), O(\3{m}_T) )$) then 
\begin{equation}
(k_A + l)^2 = \Theta\left(m Q \right) \, , \qquad (k_B - l)^2 = \Theta\left(m Q\right) \, . \label{eq:l.soft}
\end{equation}
These are far off-shell, and therefore in a relatively
  short-distance region where standard weak coupling pQCD physics
  applies.

We will now show that we can apply a contour deformation
that takes the integration out of the Glauber region, so that $l$ is at least soft and Eq.~\eqref{eq:l.soft} is fulfilled.  This is a
standard result in the theory of factorization, but the detailed
demonstration is slightly modified.  The modification takes care of
the fact that we have an array of hadrons filling in a rapidity range
with no big gaps, whereas the usual arguments in factorization theory
have configurations only with widely separated momenta.  

To exhibit the contour deformation conveniently, we write each
$k_j$ in the form $q_j+l$, where $q_j$ is the value obtained by computing $k_j$ from
the external momenta when $l=0$, i.e.,
\begin{equation}
  q_j = k_A - \sum_{i=1}^j p_i = - k_B + \sum_{i=j+1}^N p_i .
\end{equation}
We deform the contour by adding an
imaginary part to $l^{\pm}$:
\begin{equation}
\label{eq:deform}
  l^+ = l_R^+ - i \Delta(l_R), 
\qquad
  l^- = l_R^- + i \Delta(l_R), 
\end{equation}
where $l_R^{\pm}$ are the real parts.  The deformation is implemented by
increasing the real number $\Delta$ from zero to a positive value. In
general, we will allow $\Delta$ to depend on $l_R$, and we can also allow a
different imaginary part of the two components of $l$.  But initially,
for use in the Glauber region, we will assume the 2 imaginary parts to
be approximately equal and constant.
Once $l_R^{\pm}$ is well outside the Glauber region, different
  amounts of deformation are allowed than in the Glauber region.  But
  here we are only concerned with the deformation in the Glauber
  region, since that is the region in which the MPM is intended to be
  applied as a useful approximation to non-perturbative strong
  interactions.

\begin{figure}
\begin{tabular}{c}
\includegraphics[scale=0.3]{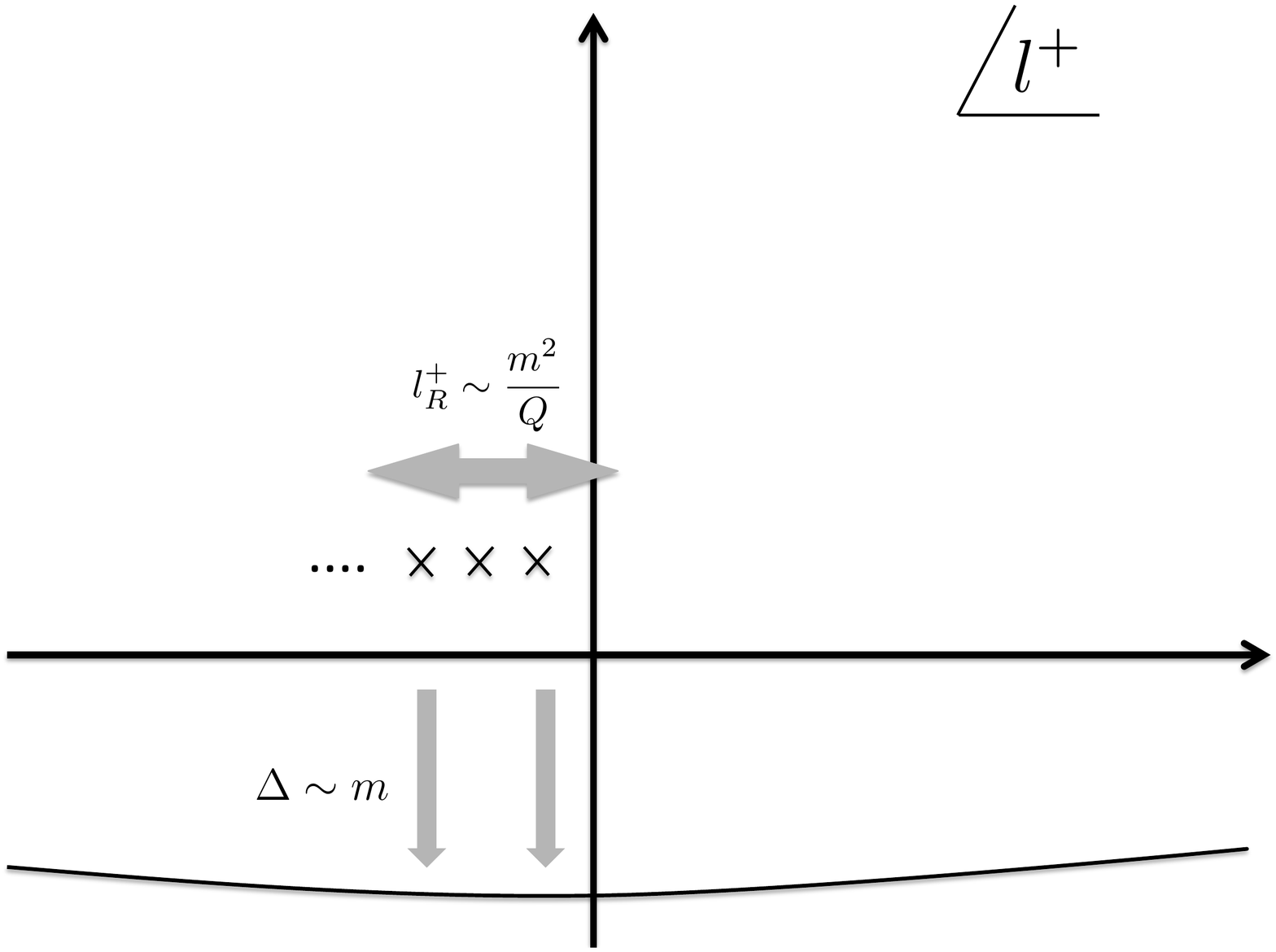}  \\
\vspace{1mm} \\
\includegraphics[scale=0.3]{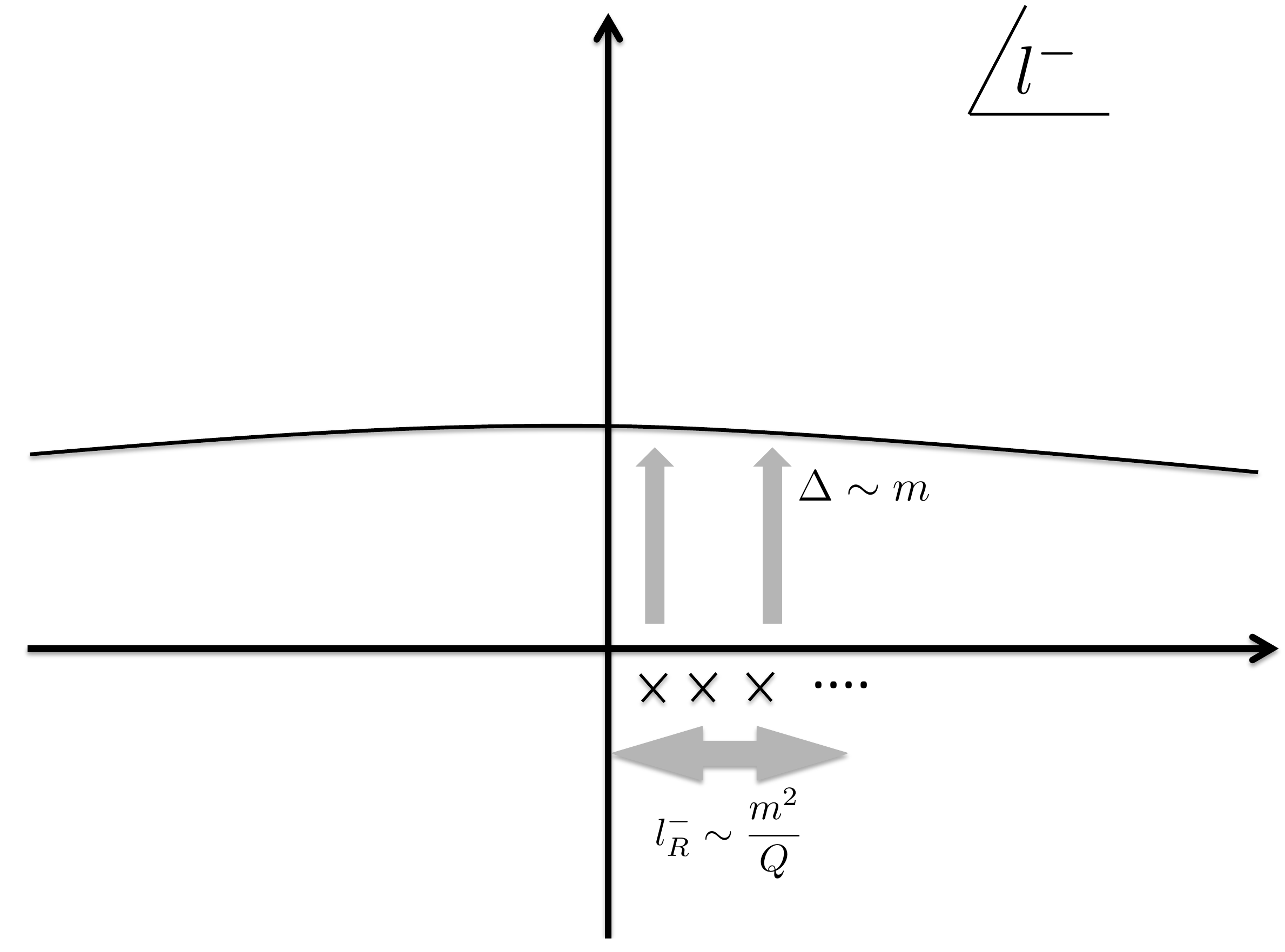} 
\end{tabular}
\caption{Singularity structure and contour deformations on
    $l^+$ and $l^-$ when $l$ is in the Glauber region ($|l^+l^-| \ll
    m^2$). On the deformed contour, some propagators are off-shell by
    order $\Theta(m Q)$---see
    Eqs.~\eqref{eq:im.denom}--\eqref{eq:im.denom.kb}.}
\label{fig:contourdef}
\end{figure}

The denominator $k_j^2-m_q^2+i\epsilon = (q_j+l)^2-m_q^2+i\epsilon$ is
\begin{equation}
  2 (q_j^+ + l^+)(q_j^- + l^-) - (\3{q}_{j,T}+ \3{l}_T)^2-m_q^2 + i\epsilon,
\end{equation}
whose real part is
\begin{equation}
\label{eq:re.denom}
  2 (q_j^+ + l_R^+)(q_j^- + l_R^-) - (\3{q}_{j,T}+ \3{l}_T)^2-m_q^2 + 2\Delta^2,
\end{equation}
and whose imaginary part is
\begin{multline}
\label{eq:im.denom}
  \epsilon + 2(q_j^+ - q_j^- + l_R^+ - l_R^-) \Delta
\\
  =
  \epsilon + 2\left[ \Theta(m\cosh(y_j)) + l_R^+ - l_R^- \right] \Delta, 
\end{multline}
given the estimates of $q_j^{\pm}$ that follow from \eqref{eq:kj+} and
\eqref{eq:kj-}.  Also, the imaginary parts of the propagator denominators $(k_A + l)^2 - m^2 + i \epsilon$ and 
$(k_B - l)^2 - m^2 + i \epsilon$ are, respectively,
\begin{align}
\label{eq:im.denom.ka}
2 \Delta \left( k_A^+ - l^+_R + l_R^- \right) + \epsilon &{}= 2 \Delta \left( \Theta(Q) - l_R^+ + l_R^- \right) \, , \\
\label{eq:im.denom.kb}
2 \Delta \left(k_B^- - l^+_R + l_R^- \right) + \epsilon &{}= 2 \Delta \left( \Theta(Q) - l_R^+ + l_R^- \right) \, . 
\end{align}

When $l_R^{\pm}$ are both smaller in size than $m$,
as in the Glauber region, the imaginary part of every denominator is
positive, so we have a successful deformation. 

In the integration over $l_R$, once $l_R^+$ or $l_R^-$ gets to be
bigger than order $m$ in size (and negative or positive,
respectively), the imaginary part no longer retains its sign. For
these larger values of $l_R^{\pm}$, a different deformation is needed;
this traps the integral over $l^{\pm}$ in a region where the two
components are of order $m$.  But this is a region far beyond where
the MPM was intended to be appropriate, for at least one quark line
goes far off-shell, with a virtuality of order $Qm$ or more.

When $l_R^{\pm}$ is smaller than $m$, and especially much smaller, the
allowed deformation enables us to get an imaginary part for $l^{\pm}$
that is of order $m$, by taking $\Delta$ of order $m$.  It therefore
follows that everywhere on the (deformed) integration contour, at
least one of the longitudinal components of $l$ is at least of order
$m$, and that we have propagators that are far off-shell, with
virtuality of order $Qm$. (See Fig.~\ref{fig:contourdef}.)

It might be naturally supposed that a model for non-perturbative
physics would have some appropriate cut offs to remove any contribution from 
far off-shell propagators.  But the cut offs should
still obey standard relativistic causal and analytic properties.
Therefore we can still deform out of the Glauber region into a
  region where the model is inappropriate.

Notice that most of the other quark lines
also go far off-shell, since on the deformed contour
$(q_j+l)^2-m_q^2 = \Theta(m^2 \cosh y_j)$.  Only the propagators for the
low rapidity lines stay at low virtuality.  Thus we get a strong
suppression of the graph, compared with the power-counting estimate
obtained from the power-counting that would be appropriate for the
Glauber region.

It follows from the above arguments that the unadorned MPM does not
adequately model the phenomena that it was intended to describe.

Note that this objection does not apply to the MPM applied to soft
hadron-hadron collisions.  To allow the contour deformation we needed
a loop momentum that circulated through the hard scattering vertex;
but a relevant loop does not exist in the case of hadron-hadron
scattering.

\section{String-like MPM}
\label{sec:string}

In reality, a gluon field is created between the outgoing
quark-antiquark pair.  To allow for this, and for the creation of
quark antiquark pairs in the gluon field, a simple model has the
structure of Fig.\ \ref{fig:mpm2}.  Here gluons are emitted from the
quark and antiquark, and then we have attached one gluon to each of what
in the MPM were quark lines joining neighboring hadrons.  
We will show that the integration is trapped in a region where
  the explicitly shown lines in Fig.\ \ref{fig:mpm2} all have
  virtuality of order $m^2$.  We will also show that in this region
  there is a radical change in the directions of the flow of longitudinal
momentum on these quark lines, compared with the simple MPM.  The quark
lines are given different arrows than in the MPM; this is a mnemonic
to indicate an important flow of positive components of momentum that
is very different than in the simple MPM.
There will remain lines that are far off shell, but these are
  in the shaded blob in Fig.\ \ref{fig:mpm2}.

To get these results, it is not necessary that exactly one gluon
attach to each quark-line segment between hadrons; our results will
apply also if multiple gluons attach to a segment, or if not too high
a proportion of the segments have no gluon.  The particular case in
Fig.\ \ref{fig:mpm2}, with one gluon per segment, simply provides one
specific case to illustrate the principles.

It should be observed that not only can the diagrammatic structure of
Fig.\ \ref{fig:mpm2} be considered as implementing the string model,
but that it is also related to a diagrammatic formulation of the
cluster-hadronization model \cite{Corcella:2000bw, Bahr:2008pv}.
Cluster hadronization is the other major hadronization model used in
Monte-Carlo event generators.

A possible set of $N$ independent loop variables are the momentum of
the upper quark, $l_0$, and the momentum $l_j$ of each of the $N-1$
gluon lines that connect to the (now modified) multiperipheral ladder.
In accordance with the general features of the hadron kinematics, we
assume that transverse momenta of these lines are of order $m$, and
that the rapidity of $l_j$ is similar to the rapidity $y_j$ of a
hadron near where it connects.  That is, each gluon connects to a part
of the ladder with similar rapidity to that of the gluon.  These
assumptions are appropriate to the non-perturbative physics we wish to
model.  We will term this region of kinematics the canonical region
for the model.  Integration outside the canonical region puts some
propagators in the ladder much further off shell than $m^2$ and
corresponds to different physics.  The important issue is now to
determine whether or not the integration is trapped in the canonical
region, and we will find that it is indeed trapped.

\begin{figure}
  \includegraphics[scale=0.5]{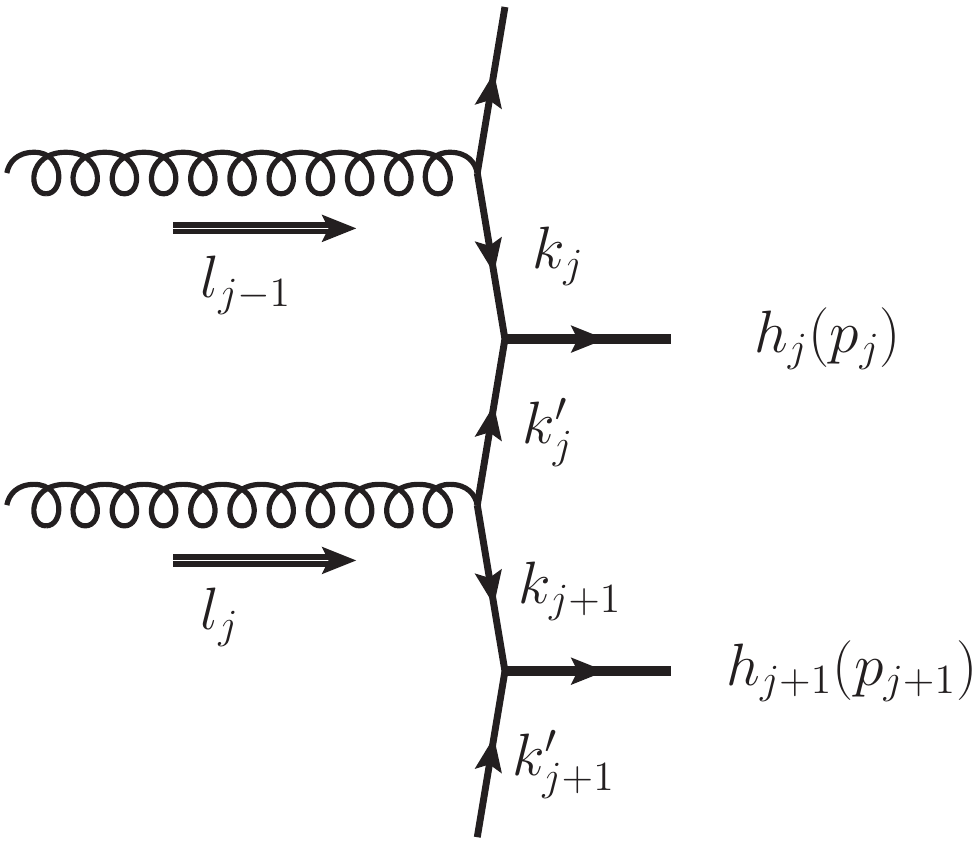}
  \caption{Component of string-like MPM.}
  \label{fig:component}
\end{figure}

We label the quark-line momenta as follows: $k_j$ is the momentum
entering the vertex for hadron $j$ from above, and $k'_j$ is the
momentum entering from below --- Fig.\ \ref{fig:component}.  By
momentum conservation, 
\begin{align}
p_j &= k_j+k'_j; \qquad &1 < j < N \\
l_j  &= k'_j+k_{j+1}; \qquad &0 < j < N \\
p_1&= l_0 + k_1' \\
p_N&= l_N + k_N \, ,
\end{align}
where 
\begin{equation}
l_N := q - \sum_{j=0}^{N-1} l_j \, .
\end{equation}

To further analyze the kinematics of the canonical region and to
  locate the conditions for the integration to be trapped there, we
  find it useful to change variables.
We define fractional momentum variables at each
quark-hadron vertex by:
\begin{equation}
  x_j = k_j^+/p_j^+; \qquad 1 <  j \leq N \, ,
\end{equation}
\begin{equation}
  y_j = {k'}_j^-/p_j^-; \qquad 1 \leq  j < N \, .
\end{equation}
(The use of $k'_j$ instead of $k_j$ in the second equation is to give
a kind of symmetry under exchange of the roles of the initial quark
and antiquark.)
We define $x_1$ and $y_N$ by writing
\begin{equation}
l_0^+ = x_1 p_1^+, \qquad l_N^- = q^- - \sum_{j=0}^{N-1} l_j^- = y_N p_N^- \, . \label{eq:l0lN}
\end{equation}
For the middle hadrons, $1 < j < N$, 
\begin{gather}
  k_j^+ = x_j p_j^+, \quad k_j^- = (1-y_j) p_j^-,
\\
  {k'}_j^+ = (1-x_j) p_j^+, \quad {k'}_j^- = y_j p_j^-,
\end{gather}
and hence
\begin{align}
  l_j^+ = {}& (1-x_j) p_j^+ + x_{j+1} p_{j+1}^+,
\\
  l_j^- = {}& y_j p_j^- + (1-y_{j+1}) p_{j+1}^- ,
\end{align}
while for $j = 1$ and $j = N$, we have 
\begin{flalign}
{k'}_1^+ &= p_1^+ - l_0^+ = (1 - x_1) p_1^+;  & {k'}_1^- &= y_1 p_1^-  \\
k_N^+ &= x_N p_N^+; & k_N^- &= (1 - y_N) p_N^- \, .
\end{flalign}
(Note that there is no $k_1$ and no $k'_N$.)

The momentum on the bottom quark line has plus component
\begin{align}
  l_N^+ = q^+-\sum_{j=0}^{N-1} l_j^+ = {} & p_N^+ (1-x_N) \, , \label{eq:bottomplus} 
\end{align}
while the momentum of the top quark line has minus component
\begin{align}
  l_0^- =  p_1^- (1-y_1) \label{eq:topminus} \, .
\end{align}

The integration over the longitudinal components of the $l_j$ can be
changed to integration over $x_j$ and $y_j$ with a simple Jacobian:
\begin{align}
  \prod_{j=0}^{N-1} ( dl_j^+ dl_j^- )
  = {}&
  \prod_{j=1}^N ( p_j^+ p_j^- ) \prod_{j=1}^N ( dx_j dy_j )
\nonumber\\
  = {}&
  \prod_{j=1}^N \left( \frac{E_{j,T}^2}{2} \right) \prod_{j=1}^N ( dx_j dy_j ).
\end{align}

As in the previous section, we use the term ``canonical region" to refer 
to the momentum region that is consistent with the spacetime picture of hadronization.
Thus, in the canonical region, Eqs.~\eqref{eq:l0lN}--\eqref{eq:topminus} mean that all the transverse momenta are of order $m$
and the $x_j$ and $y_j$ variables of order unity, and hence all the
propagator denominators are of order $m^2$.  

What we would mean if the integration were not trapped would be the following: There would be an
allowed contour deformation such that everywhere on the contour at
least one of the quark lines shown in Fig.\ \ref{fig:mpm2} is much
further off-shell than order $m^2$.  We will find that in fact no such
deformation is possible.

To specify an allowed deformation, we parameterize the surface of
integration by the real parts of the values of the integration
variables.  We restrict our attention to the longitudinal
  momentum components, and work in terms of the fractional momenta
  $x_j$ and $y_j$.
We use a real value $\lambda$ taking values in the range $0\leq \lambda \leq
1$ to parameterize the deformation starting from real momenta.  
We therefore write
\begin{align}
  x_j &= x_{R,j} + i \lambda x_{I,j}(x_R,y_R) ,
\\
  y_j &= y_{R,j} + i \lambda y_{I,j}(x_R,y_R) .
\end{align}

By Cauchy's theorem (generalized to multiple variables), the value of
the integral is independent of $\lambda$ provided that the deformation is
allowed, i.e., that no poles are crossed by the contour when $\lambda$
is increased from 0 to 1. There must be
inserted in the integral a factor of the Jacobian for the
  transformation of variables from $x_{R,j}$ and $y_{R,j}$ to the
  complex variables $x_j$ and $y_j$.

Let us consider the region of integration where $x_R$ and $y_R$ are
between $0$ and $1$.  Then, from
Eqs.~\eqref{eq:l0lN}--\eqref{eq:topminus}, the contour of
integration passes through the canonical region if $\lambda = 0$.
We will show that the integral is trapped in this region, and
  to do this we must show that there exists no choice of 
  $x_{I,j}(x_R,y_R)$ and $y_{I,j}(x_R,y_R)$ such that we can
  increase $\lambda$ from 0 to 1, without crossing any poles, and such that
  the deformations obey
  $x_{I,j}\gg1$ and $y_{I,j}\gg1$ for at least some values of
  $j$.  If such a deformation were to exist it would take the
  integration outside the canonical region.  We will also require that
  the sizes of the derivatives, $|\partial x_{I,j}/ \partial x_{R,k}|$ etc all stay
  bounded, say below 1, so that strongly varying structures in the imaginary
  parts don't exist, and the Jacobian remains of order unity.

  There do indeed exist non-trivial allowed deformations, but these
  all have $x_{I,j}$ and $y_{I,j}$ of order unity at most, and
  therefore stay in the canonical region.

Consider the relevant denominators:
\begin{align}
\label{eq:kj.denom}
  k_j^2-m_q^2+i\epsilon
  ={}& 2 x_jp_j^+ (1-y_j) p_j^- - k_{j,T}^2 - m_q^2 + i\epsilon
\nonumber\\
  ={}& E_{j,T}^2 x_j(1-y_j) - k_{j,T}^2 - m_q^2 + i\epsilon ,
\\
\label{eq:kjprime.denom}
  {k'}_j^2-m_q^2+i\epsilon
  ={}& 2 (1-x_j)p_j^+ y_j p_j^- - {k'}_{j,T}^2 - m_q^2 + i\epsilon 
\nonumber\\
  ={}& E_{j,T}^2 (1-x_j)y_j - {k'}_{j,T}^2 - m_q^2 + i\epsilon ,
\\
\label{eq:lj.denom}
  l_j^2-m_g^2+i\epsilon
  ={}& 2 [(1-x_j)p_j^+ + x_{j+1}p_{j+1}^+] \times
\nonumber\\ &  \hspace*{-2cm}
        \times [y_jp_j^- + (1-y_{j+1})p_{j+1}^-]
       - l_{j,T}^2 - m_g^2 + i\epsilon.
\end{align}
(Here, $m_g$ is a mass scale of order $\Lambda_{\rm QCD}^2$ representing the 
effects of confinement in cutting off soft gluons.)
The transverse momenta and masses are of order $m$, as are the
products of $p_j^+$ and $p_{j+1}^+$ with $p_j^-$ and $p_{j+1}^-$.  

If we could find an allowed deformation out of the canonical
  region, at least one of the denominators would need to be much
  larger than $m^2$ on the deformed contour.  As mentioned
  above, we are considering the part of the integration region where
the real parts $x_{R,j}$ and $y_{R,j}$ are between zero and one.
Then
all of $x_{R,j}$, $1-x_{R,j}$, $y_{R,j}$, and $1-y_{R,j}$ are
positive.  Both of the momenta $k_j$ and $k'_j$ are therefore
future-pointing as regards both their (real) plus- and
minus-components, which is unlike corresponding momenta in the pure
MPM of Fig.\ \ref{fig:mpm}.

Furthermore, to give an easy demonstration the non-existence
  of this hypothesized deformation, we restrict to the case that all
  of $x_{R,j}$, $1-x_{R,j}$, $y_{R,j}$, and $1-y_{R,j}$ are order
  unity rather than some being much less than unity.

The large size of $x_j$ or $y_j$ is
achieved by the imaginary part, i.e., $|x_{I,j}|\gg1$ and/or
$|y_{I,j}|\gg1$. 
The imaginary parts of the denominators for $k_j$ and $k'_j$ are
\begin{align}
\label{eq:kj.I}
  & \Im(k_j^2-m_q^2+i\epsilon)
\nonumber\\  & \hspace{1cm}
  = \lambda E_{j,T}^2 [ x_{I,j} (1-y_{R,j}) - y_{I,j} x_{R,j} ] + \epsilon ,
\\
\label{eq:kj.prime.I}
  & \Im({k'}_j^2-m_q^2+i\epsilon)
\nonumber\\  & \hspace{1cm}
  = \lambda E_{j,T}^2 [ - x_{I,j} y_{R,j} + y_{I,j} (1-x_{R,j}) ] + \epsilon .
\end{align}
Since $\epsilon > 0$, then, starting at $\lambda=0$, to be able to deform off the real axis without crossing a pole the
coefficients of $\lambda$ must be positive when the real parts of the momenta
are at the corresponding pole.  At the pole on the $k_j$ propagator,
with $\lambda=0$, we have $1-y_{R,j} = (k_{j,T}^2+m_q^2)/(E_{j,T}^2
x_{R,j})$.  So positivity of the imaginary part of the propagator for
$k_j$, as we deform the contour off the real axis requires that
\begin{equation}
  (k_{j,T}^2+m_q^2) \frac{x_{I,j}}{x_{R,j}} - E_{j,T}^2 x_{R,j} y_{I,j} 
\end{equation}
be positive for all $x_{R,j}$ between zero and one.  Hence $x_{I,j}>0$
and $y_{I,j}<0$. But exactly the opposite condition applies to get a
positive imaginary part for the denominator for the other line,
$k'_j$. That is, at the pole for $k'_j$ with $\lambda = 0$, we have 
$1-x_{R,j} = ({k'}_{j,T}^2+m_q^2)/(E_{j,T}^2 y_{R,j})$. The condition for doing a deformation without 
crossing a pole is 
\begin{equation}
  ({k'}_{j,T}^2+m_q^2) \frac{y_{I,j}}{y_{R,j}} - E_{j,T}^2 y_{R,j} x_{I,j} > 0\, .
\end{equation} 
This requires $x_{I,j}<0$ and $y_{I,j}>0$. Thus, allowed deformations for $k_j$ lines clash with allowed deformations for $k'_j$ lines.
So no deformation far out of the canonical region is possible
for any set of $x_{I,j}$ or $y_{I,j}$.

The poles on the two lines $k_j$ and ${k'}_j$ are at different
  locations in $x_j$ and $y_j$.  We could therefore imagine avoiding
  them by changing the sign of the imaginary part of one or more
  integration variables between the poles.  But since to get a large
  denominator the imaginary part of the variable has to be much larger
  than unity, the derivative with respect to the real part would also
  be much larger than unity, which violates the smoothness requirement
  adopted earlier.

We have found that examination of the $k_j$ and $k'_j$ denominators is
sufficient to show that the integration is trapped in the canonical
region, with the denominators being of order $m^2$.  In this region,
the denominators for the gluon lines $l_j$ are also of order $m^2$.
In fact these denominators also participate in the trapping.  For
example, if we have the $x_{I,j}$ positive and the $y_{I,j}$ negative
to avoid the poles of the $k_j$ lines, as above, then the imaginary
part of the $l_j$ denominator (\ref{eq:lj.denom}) does not have a
fixed sign, and so the hypothesized deformation is not allowed.

Notice that getting a situation where the contour is trapped
  in the canonical region depends on the gluons being able to inject
  appropriate amounts of momentum into the would-be multiperipheral
  ladder.  Then about half the plus- and minus- momentum components on
  the quark lines are reversed in sign compared with the elementary
  MPM.

\section{Discussion}
\label{sec:discuss}

We have shown that in the elementary MPM for non-perturbative
hadronization in $e^+e^-$ annihilation, the loop integration can be
deformed far out of the momentum region appropriate to its
hypothesized validity.  A minimal requirement for an appropriate
non-perturbative model of a Feynman-graph kind, is that it incorporate
the effects of gluon emission in a string-like way, as in Fig.\
\ref{fig:mpm2}.

There are dramatic differences in the directions of momentum flow and
in the space-time structure between that graph and the elementary MPM
Fig.\ \ref{fig:mpm}.  In the MPM, quark lines, like $k_A+l$, are far
off-shell after the contour deformation.  So the fastest hadrons are
formed first.  This is in complete contrast \cite{Casher:1974vf} to
string-like models, where the fastest hadrons are formed last, on an
appropriate time-dilated scale.

The importance of these results is for the formulation and
interpretation of models of non-perturbative hadronization of hard
jets.  For example, the successful string-inspired model of Refs.\
\cite{Kerbizi:2017rpp, Artru:2012zz, Artru:2010st,
  Artru:in2p3-00953539} is formulated in terms of multiperipheral
graphs.  The use of a Feynman graph formulation allows the systematic
and consistent incorporation of spin effects.  This is in contrast to
purely semi-classical formulations of a string hadronization, where
the use of the ideas of classical, non-quantum physics makes it much
harder to see how to incorporate the intrinsically quantum mechanical
phenomenon of quark spin.  The results in the present paper indicate
that the graphs must be interpreted as containing
momentum-non-conserving vertices or propagators in an external gluon
field.  The consequences can be seen in Ref.\ \cite{Kerbizi:2018qpp},
where there is an explicit allowance for transfer of energy and
momentum between the quarks and the string, with a non-trivial
identification of the momentum to use in the vertices of the
multi-peripheral graph.

The results in Ref.\ \cite{Matevosyan:2016fwi, Bentz:2016rav} are
developed from earlier work \cite{Ito:2009zc} that used a chiral model
of the Nambu-Jona-Lasinio (NJL) \cite{Nambu:1961tp, Nambu:1961fr} type
applied to a lowest order graph for fragmentation of a quark into a
pion plus a quark.  After iteration to apply to multiple splittings,
such a model gives graphical structures like that of the MPM.
Calculations such as those in \cite{Ito:2009zc} applied to a splitting
$q_j\to\pi_j+q_{j+1}$, have the final quark on-shell, and the initial
quark off-shell.  But to iterate \cite{Matevosyan:2016fwi} the
splitting, the final quark, $q_{j+1}$ is changed to be off-shell.
This is formulated according to the Field-Feynman structure; it
amounts to a kind of momentum non-conservation appropriate to
implementing a string model. Another useful resources is Ref.~\cite{Accardi:2009qv}, which 
addresses the spacetime structure of hadronization in  the Lund string model in Sec.\ 2.4. 
Such methods might help further clarify the relationship between Feynman graph 
structures and hadronization.

Chiral models can be argued to be useful as effective low-energy
approximations to QCD.  These models are important because they aim to
capture the properties of QCD associated with chiral symmetry
breaking.  The results in this paper suggest an important direction
for enhancing models such as the one in Ref.\ \cite{Schweitzer:2012hh}
to apply to high-energy dynamical processes like the non-perturbative
hadronization of hard jets.  This is to formulate the theories to
apply to processes that occur in the background of an non-vacuum state
that corresponds to a gluonic flux-tube.  There are probably some
useful tools to relate strings and field theory in the paper by
\citet{Artru:1986vc} on the quantization of the string by a
sum-over-histories method.

\begin{acknowledgments}	
  This work was supported in part by the
  U.S. Department of Energy under Grant No.\ DE-SC0013699. 
  T.~Rogers's work was supported by the U.S. Department of Energy, Office of Science, 
  Office of Nuclear Physics, under Award Number DE-SC0018106.
  This work was also supported by the DOE Contract No. DE- AC05-06OR23177, under which Jefferson Science Associates, LLC operates Jefferson Lab.
  We acknowledge useful discussions with M.~Diefenthaler and with H.~Matevosyan. 
\end{acknowledgments}

\appendix

\section{Estimate of number of particles per unit rapidity}
\label{sec:TASSO}

In Ref.\ \cite{Braunschweig:1990yd}, the TASSO experiment reported
results on the distribution of charged hadrons in jets in $e^-e^-$
annihilation at $Q$ between 14 and 44 GeV.  In Table 9 are shown
values for the normalized cross section
$(1/\sigma_{\text{tot}})d\sigma/d\ln(1/x)$.  Here $x=2p/Q$, where $p$ is the
momentum of the detected particle; to a leading approximation, $x$
corresponds to the fragmentation variable $z$.  We now roughly extract
from this data the number of hadrons per unit rapidity in a jet. 

For hadrons of high rapidity, Eq.\ (\ref{eq:pj}) gives a corresponding
$x$ value:
\begin{equation}
  x = \frac{E_{j,T}}{Q} e^{y_j}.
\end{equation}
Hence the rapidity distribution in fragmentation is given by $dN/dy =
(1/2\sigma_{\text{tot}})d\sigma/d\ln(1/x)$.  The extra factor of $1/2$ is to
compensate the fact that the $x$ distribution gets a contribution from
each jet.  The most common particles in jets are pions. Isospin and
charge conjugation symmetry show that the neutral pion fragmentation
function in a $u$ or $d$ quark is half the charged pion fragmentation
function: $f_{\pi^0/u} = (f_{\pi^+/u}+f_{\pi^-/u})/2$, etc.  Thus the total
hadronic number distribution is $3/2$ times the charged hadron
distribution. Hence $dN^{\text{all}}/dy =
(1/2\sigma_{\text{tot}})d\sigma^{\text{ch.}}/d\ln(1/x)$.  In Table 9 of Ref.\
\cite{Braunschweig:1990yd}, the values of this quantity first increase
as $\ln(1/x)$ is increased from 0 (which corresponds to $x=1$ or
$z=1$).  They reach a peak and then decrease.  Now, once the rapidity
is lower than a unit or two, it is not appropriate to apply the above
approximations. So we take the peak value as the relevant one.
Furthermore, as $Q$ increases, perturbative gluon emission becomes
more important; this increases the hadron multiplicity beyond what it
appears appropriate to attribute to purely non-perturbative QCD.  So we
take the peak in $(1/\sigma_{\text{tot}})d\sigma/d\ln(1/x)$ at
$Q=\unit[14]{GeV}$ as the relevant one to estimate $dN/dy \simeq 3$.  This
is the number of hadrons per unit rapidity in the interior of the
string.

\bibliography{jcc}

\end{document}